\documentclass{aa}
\usepackage{graphicx}
\usepackage{txfonts}
\usepackage[colorlinks=true,
            urlcolor=blue,
            linkcolor=blue,
            citecolor=blue]
            {hyperref}
\begin{document} 

   \title{Evidence for gravitational self-lensing of the central supermassive black hole binary in the Seyfert galaxy NGC 1566}

      \author{W.~Kollatschny \inst{1},
              D.~Chelouche \inst{2,3}
              }

   \institute{
          Institut f\"ur Astrophysik und Geophysik, Universit\"at G\"ottingen,
          Friedrich-Hund Platz 1, 37077 G\"ottingen, Germany \\
          \email{wkollat@gwdg.de}
          \and
          Department of Physics, Faculty of Natural Sciences, University of Haifa, Haifa 3498838, Israel
          \and
          Haifa Research Center for Theoretical Physics and Astrophysics, University of Haifa, Haifa 3498838, Israel
          }

  \date{Received 16 August 2024 / Accepted 11 September 2024}

 \abstract
 {}
{It is generally accepted that all massive galaxies host supermassive black holes (BHs) in their center and that mergers of two galaxies lead to the formation of BH binaries. The most interesting among them comprise the mergers in their final state, that is to say\ with parsec (3.2 light years) or sub-parsec orbital separations.  It is possible to detect these systems with binary self-lensing. }
{Here we report the potential detection of a central supermassive BH binary in the active galaxy (AGN) NGC\,1566 based on a microlensing outburst.  The light curve of the outburst -- based on observations with the All Sky Automated Survey for SuperNovae -- lasted from the beginning of 2017 until the beginning of 2020. The steep symmetric light curve as well as its shape look very different with respect to normal random variations in AGN.}
{However, the observations could be easily reproduced with a best-fit standard microlensing light curve. Based on the light curve, we derived a characteristic timescale of 155 days. During the outburst, the continuum as well as the broad line intensities varied; however, the narrow emission lines did not. This is an indication that the lensing object orbits the AGN nucleus between the broad line region (BLR) and the narrow line region (NLR), that is, at a distance on the order of 250 light days. 
The light curve can be reproduced by a lens with a BH mass of $5\times10^{5} M_{\odot}$. This implies a mass ratio to the central AGN on the order of 1 
to 10.}
{} 
% 5 {} token are mandatory


  \titlerunning{Lensing in NGC\,1566}
  \authorrunning{W. Kollatschny et al.}

  \maketitle
%
%
%**********************************************************************************
%
\section{Introduction}\label{sec:introduction}
%
%**********************************************************************************
It is generally accepted that supermassive black holes (SMBHs) exist in the center of almost every galaxy \citep[]{kormendy13}. If these galaxies interact and finally merge,  massive black hole binaries (MBHBs) are expected to be formed as a result of mergers of their host galaxies \citep[]{begelman80}.
In addition, scenarios have been proposed in which one of these nuclei might be an active galactic nucleus (AGN). Specifically, the broad emission lines can originate around both nuclei or around the more massive BH if the nuclei are in a rather final state with separations only in the sub-parsec range \citep[]{begelman80}. 
In contrast to the broad lines, the narrow lines originate at much larger distances with typical distance values on the order of more than 10 light years from the central binary BH. 

The search for and investigation of double-nucleus galaxies dates back to 1978 \citep[]{petrosyan78, netzer87}. Dozens of double nucleus galaxies were detected that were spatially resolved on direct images at that time in the optical wavelength regime. Typical distances of these nuclei are a few arcsec and correspond to projected distances of a few kiloparsec.   
The search for close MBHBs, in which one or both nuclei are active, has been an active field of research ever since. 

Supermassive black hole binaries with sub-parsec separations are of particular interest because they are in a later state of evolution. 
 However, it is difficult to spatially resolve such objects.  Two prominent cases with sub-parsec separations are the double and triple nucleus galaxies NGC\,6240 \citep[]{komossa03, kollatschny20} and OJ287 \citep[]{valtonen08, komossa23}. 
A detailed review on the quest for dual and binary SMBHs from the perspective of a multi-messenger view
has been presented by \cite{derosa19}.
A more recent review on observational signatures of SMBH binaries is given by \cite{dorazio23}.

%**********************************************************************************
%
\section{Observations of the continuum light curve of NGC\,1566} \label{sec:observations}
%
%**********************************************************************************

NGC\,1566 is a nearby face-on SABbc spiral galaxy \citep[]{devaucouleurs73}.
It has been known since the 1960s \citep[]{pastoriza70, alloin85} that NGC\,1566 is a variable Seyfert 1 galaxy. 
The BH mass of the nucleus in NGC\,1566 has previously been estimated to be  $5\times10^{6} M_{\odot}$ based on the full width at half maximum  (FWHM) of the broad Br\,$\gamma$ line as well as other optical broad emission lines \citep[]{smajic15}. Measurements of the stellar velocity dispersion 
%of $105.\pm10.$\,km\,s$^{-1}$ 
of the stars in the bulge come to a similar result with some scatter \citep[]{smajic15, dasilva17}. A total central BH mass of $5\times10^{6} M_{\odot}$ has been estimated based on CaII triplet absorption \citep[]{ochmann24}.

 A strong increase in the X-ray flux of NGC\,1566 -- in comparison to archival X-ray data from December 2004 to August 2013 taken with the BAT instrument on Swift -- was discovered by chance \citep[]{ducci18} in an observation taken with the INTEGRAL satellite from June 12 to 19, 2018.
Subsequent Swift observations \citep[]{ferrigno18} on June 26, 2018, confirmed a strong increase in the X-ray by a factor of $\sim$ 15 compared to archival data.
Further Swift observations \citep[]{grupe18} from July 10, 17, 24, and 31 verified  the high X-ray and UV flux states, but indicated a decrease with respect to the observation taken on July 10.
Follow-up observations in the X-ray with Swift and XMM \citep[]{parker19} as well as in the optical \citep[]{oknyansky19, oknyansky20, ochmann24} also confirm a decline in the optical, UV, and X-ray emission over the following two years. NGC\,1566 was still found in a low state in all Swift bands in late 2023 \citep[]{xu24}.
The optical spectra indicated a normal decrease in the continuum and the broad line fluxes, while the narrow [OIII]\,$\lambda$5007 line remained constant.
Even so, the optical long-term light curve of NGC\,1566 from 2015 to 2021 -- including the flux increase for the years 2016 until July 2018 -- has not been investigated yet.

\begin{figure}
\centering
\includegraphics[width=9.cm,angle=0]
{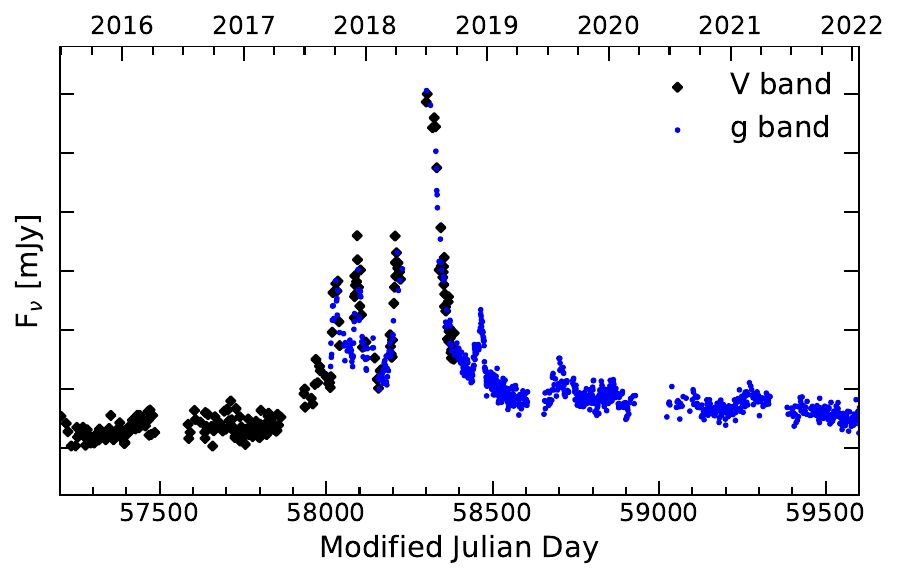}
%{V_band_lightcurve_NGC1566_lensing_20240216.pdf}
 \vspace*{-2mm} 
\caption{Combined V-band (black) and g-band (blue) light curve of the central region of NGC\,1566 for 2015 until 2023.}
\label{Ochm_ASAS_LC_2015-2022_20240216.pdf}
\end{figure}

Here we explore the optical light curve of NGC\,1566 in more detail based on  All Sky Automated Survey for SuperNovae (ASAS-SN) \citep[]{henden12} photometry from 2015 to 2022. 
ASAS-SN is a project designed to monitor the entire extragalactic sky with a mean cadence of $2-3$\,days in order to search  for transient events \citep[]{shappee14, kochanek17, jayasinghe18}. Our long-term light curve of NGC\,1566 from June 30,  2015 (MJD 57203.44) until  January 20, 2022 (MJD 59600.) is shown in
Fig.~\ref{Ochm_ASAS_LC_2015-2022_20240216.pdf}. The observational filter setup changed during the outburst, with V-band observations lasting until  September 24, 2018 (MJD 58385.23) and g-band observations starting on  September 17, 2017 (MJD 58013.36). 
This results in a considerable overlap of 372 days during which NGC\,1566 was observed in both filters. The mean and median cadence are 2.1 and 0.996 days, respectively. Due to the different filters, the V-band and g-band light curves show a shift in flux with respect to each other. In order to create a combined V-band and g-band light curve, we shifted the g-band light curve by 23.15 mJy, such that it matches the V-band light curve in the overlapping region. 
The gaps in the light curve are caused by the annual constraints due to the proximity of NGC\,1566 to the Sun.

%**********************************************************************************
%
\section{Results}\label{sec:results}
%
%**********************************************************************************

\subsection{ Shape of the continuum light curve of NGC\,1566}

\begin{figure}
\centering
\includegraphics[width=9. cm,angle=0]{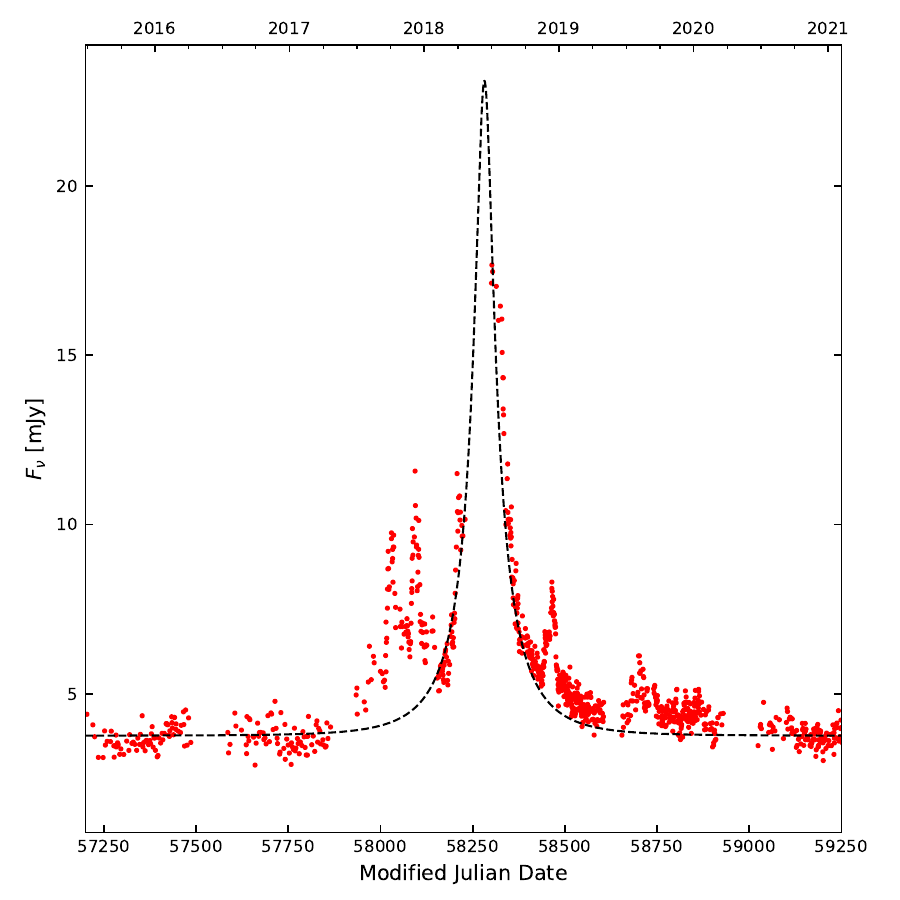}
 \vspace*{-2mm} 
\caption{Shown is the combined V-band and g-band light curve of NGC\,1566 (red) for 2015 until 2021.  
The dashed line (black) is a modeled best-fit micro-lens light curve.}
\label{Malte_lensing_fit_quad_single_20240216.pdf}
\end{figure}

\begin{figure}
\centering
\includegraphics[width=9. cm,angle=0]{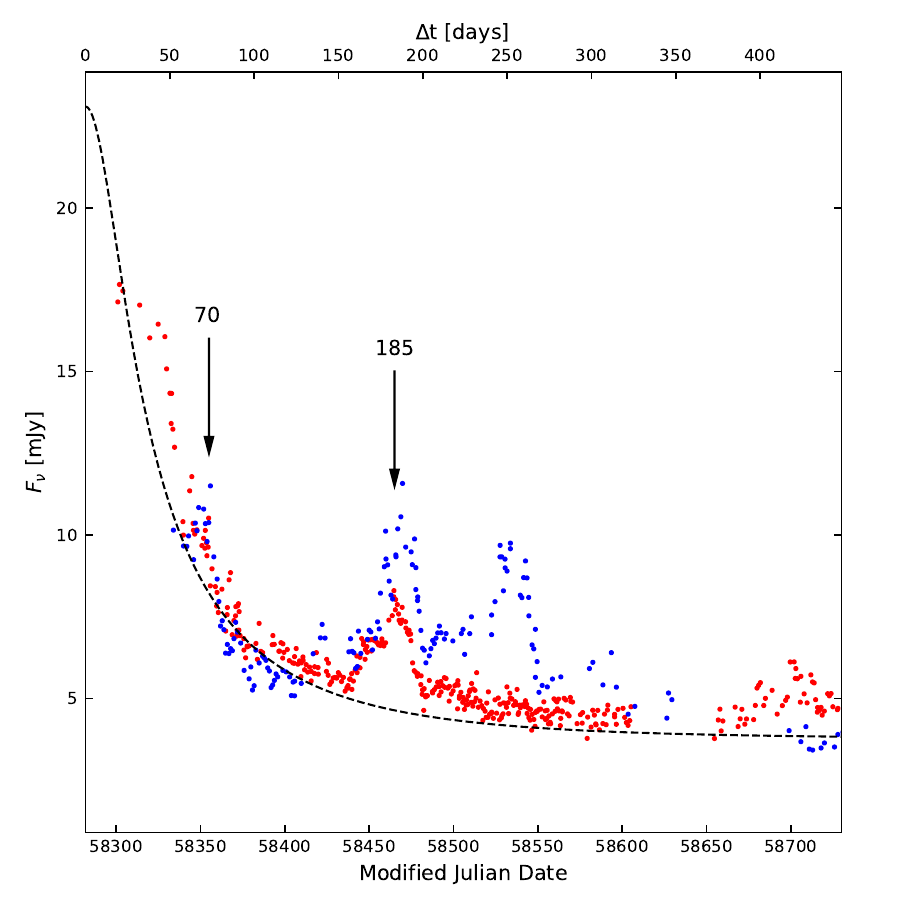}
 \vspace*{-2mm} 
\caption{Original and flipped light curve for 2018 June 12 (blue: before outburst; red: after outburst). This zoom in of Fig.~\ref{Malte_lensing_fit_quad_single_20240216.pdf} highlights the outbursts at $\pm$70 days and $\pm$185 days before and after the adopted maximum at MJD 58282.}
\label{Maltelensing_fit_zoom_20240216.pdf}
\end{figure}

The general shape of the outburst light curve looks surprisingly symmetrical with respect to its highest flux peak in mid-2018  -- with additional smaller peaks superimposed on either wing.
Fig.~\ref{Malte_lensing_fit_quad_single_20240216.pdf} shows the combined V-band and g-band light curve of NGC\,1566 (in red) from 2015 until 2021,  
%(same as Fig.~\ref{Ochm_ASAS_LC_2015-2022_20240216.pdf})
both after subtraction of %Fig.~\ref{Malte_lensing_fit_quad_single_20240216.pdf}
the respective host galaxy continuum component.
%of 62 mJy. 
This continuum component of $\sim$62 mJy is based on the modeling of the observed spectra \citep[]{ochmann24}.
% Overlaid is a flipped light curve with respect to the date 2018 June 12 
%(blue). 
The dashed line (black) is a modeled best-fit micro-lens light curve.
To test the symmetry, we inverted the observed light curve with respect to June 12, 
2018, and overlaid both the inverted and the original light curve in Fig.~\ref{Maltelensing_fit_zoom_20240216.pdf}. 
This figure is a zoom-in of Fig.~\ref{Malte_lensing_fit_quad_single_20240216.pdf}, which highlights the outbursts.

\subsection{ Explaining strong AGN variations without gravitational lensing}

Normal AGNs show stochastic flux variations \citep[]{smith18, kelly09, bischoff99, peterson02, chen23, neustadt23}.
\cite{smith18} presented a comprehensive analysis of 21
light curves of type 1 AGNs taken with the
Kepler satellite. They determined that AGNs vary substantially in terms of timescales and amplitudes. 
Furthermore, they found that the high-frequency power spectral
density function (PSD) slopes are largely inconsistent with the
values by damped random walk models. These timescales were
consistent with orbital timescales or free-fall accretion 
timescales.

\cite{macleod16} carried out a systematic search for changing-look AGNs based on 
repeated photometry from the Sloan Digital Sky Survey (SDSS) and 
Pan-STARRS1, along with repeated spectra from SDSS and the SDSS-III 
Baryon Oscillation Spectroscopic Survey. Finally they selected ten objects with large photometric variations ( > 1 mag) in their light curves and, in addition, strong broad emission line changes out of a sample
of 1011 objects. These objects have highly variable and/or 
"changing-look" broad emission line features. \citet[]{macleod16} explain the observed variations due to accretion rate changes.
\cite{graham20} searched for highly variable AGNs 
in the Catalina Real-time Transient Survey in combination with the Million Quasars (MQ) catalogue.
They searched for objects with photometric observations of more 
than one magnitude and H$\beta$/[O III] line ratio changes by >30 percent. Finally, they identified 111 sources. They explain these
observed variations with accretion rate changes. They state that the timescales of their variations imply cooling and heating fronts propagating through the disk.

Detailed investigations of selected changing-look AGNs such as
He 1136-2304, 1ES 1927+654, or Mrk1018 have recently been published
\citep[]{kollatschny18, trakhtenbrot19, brogan23}.
Strong changes in the accretion rate of the SMBH are indicated 
as the cause for the changing-look phenomenon in a recent review on changing-look 
AGNs \citep[]{ricci23}.

However, such a symmetric light curve with respect to its maximum as seen in 
NGC\,1566 is difficult to explain by means of normal AGN variability with
variable gas inflow onto the central BH:
Possible tidal disruption events (TDEs) show a steep increase 
followed by an exponential decrease \citep[]{vanvelzen21} of the continuum flux.
The light curves of rare quasi periodic eruptions (QPEs) are 
symmetric. However, their amplitude is too steep to reproduce the 
observed light curve spanning over many years \citep[]{ingram21, miniutti19}.
Furthermore, the symmetrical occurrence of the additional small
outbursts in NGC\,1566 with respect to the maximum cannot be explained by AGN variability due to variable gas inflow. 

\begin{figure}
\centering
\includegraphics[width=8.8 cm,angle=0] {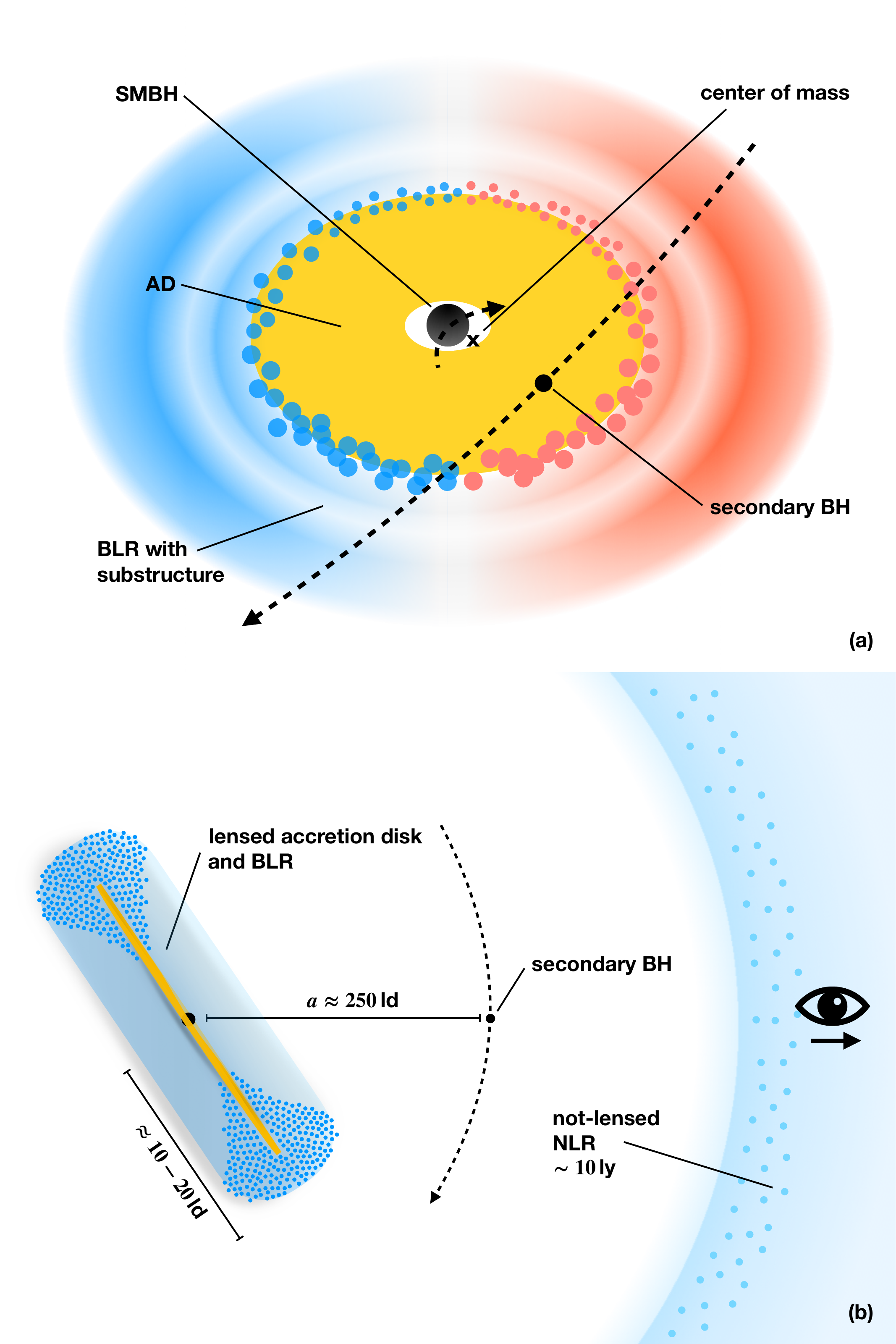}
 \vspace*{-2mm} 
\caption{Sketch of the path of the secondary lensing SMBH that is orbiting the center of mass together with the central SMBH and its BLR in NGC\,1566. We made the assumption that one half of the emission is redshifted or rather blueshifted as observed in 3C\,273 \citep[]{gravity18}. }
\label{NGC1566_lensing_v2_stackedMalte.pdf}
\end{figure}

\subsection{Modeling the observed outburst in NGC\,1566 with gravitational lensing}

However, the observed light curve of NGC\,1566 can be reproduced immediately with a simple microlensing model. 
The light curve of a microlensing event is symmetric around the brightness maximum and has a characteristic shape \citep[]{paczynski91}. This distinguishes a microlensing event from other types of variability. 
The dashed black line in Fig.~\ref{Malte_lensing_fit_quad_single_20240216.pdf} represents the best fit for the single-lens microlensing model (see Appendix). 
Our best fitting parameters are the following: (1) the time $t_{max}=$ MJD 58282 (June 12, 2018); (2) the amplitude $A = 6.3$  of the maximum magnification, after correction for the host galaxy; and (3) the characteristic timescale $t_0 = 155\pm20\,$days of the lensing event.

The derived characteristic timescale $t_0$ of the variability contains the combined information on the lens mass, the distances to the lens and source, and the transverse velocity \citep[]{paczynski96}. The mass of the lens can be estimated if reasonable assumptions can be made as to the distances and transverse velocities.
The intensity of the [OIII]\,$\lambda$5007 emission remained constant \citep[]{oknyansky20, ochmann24} from 2015 to 2021.
However, the host-corrected continuum flux 
as well as the intensities of the broad emission lines varied. 
This means that the lensing object must be located between the outer broad line region (BLR) (10 to 20 light days) and the inner narrow line region (NLR) (a few light years). Therefore, as a first-order approximation, we assume  that the secondary lensing object is located at a distance of 250 light days from the central nucleus. 
A sketch of the assumed geometry of the BLR surrounding the central SMBH with its accretion disk (AD) and the outer NLR (not to scale) is presented in Fig.~\ref{NGC1566_lensing_v2_stackedMalte.pdf}.
Furthermore, the path of the secondary lensing SMBH is indicated.

For a distance of 250 light days for the BHs, taking a combined central mass $M_\text{BH}$ of $5\times10^{6} M_{\odot}$, and assuming Keplerian orbits, we get a value of 320\,km\,s$^{-1}$  for the transverse velocity. A velocity on this order of magnitude can be expected in the intermediate region between the BLR and NLR.  
The velocities in the broad and narrow line regions of AGNs are based on their emission line widths.
Finally, for this scenario, we derived a BH mass $M_\text{BH}$ of $5\times10^{5} M_{\odot}$ for the lensing object in NGC\,1566 based on
the observed characteristic timescale of the light curve (155 days), on the distance of the lens to the source
(250 light days), and on the transverse velocity (320\,km\,s$^{-1}$). Further, we derived an orbiting time for the double nucleus system of about 4000 years assuming Keplerian orbits.

We calculated the transverse velocity as well as the masses of the lensing and the lensed BH for two additional scenarios with distances of 160 and 400 light days, respectively. The parameters are given in Table~\ref{tab:spectroscopy_log}.
\begin{table}[ht!]
\caption{Distances $d$, transverse velocities $v$, $m_{\text{lens}}$, and $M_{\text{center}}$ for three possible lensing scenarios, and a combined mass of the two BHs $M_{\text{center}}+m_{\text{lens}} \approx 5\times10^6\,M_{\odot}$.}
\centering
 \resizebox{0.49\textwidth}{!}{
    \begin{tabular}{lccc}
        \hline \hline
        \noalign{\smallskip}
          & Scenario 1 & Scenario 2 & Scenario 3 \\
        \noalign{\smallskip}
        \hline 
        \noalign{\smallskip}
        $d$\,[ld]               & 160 & 250 & 400 \\
        $v$\,[km\,s$^{-1}$]     & 400 & 320 & 255 \\
        $m_{\text{lens}}$              & $1.2\times10^{6}M_{\odot}$ & $4.8\,\,\,\,\times10^{5}M_{\odot}$ & $1.9\,\,\,\,\times10^{5}M_{\odot}$ \\
        $M_{\text{center}}$            & $3.8\times10^{6}M_{\odot}$ & $4.5\,\,\,\,\times10^{6}M_{\odot}$ & $4.86\times10^{6}M_{\odot}$ \\
        $M_{\text{center}}+m_{\text{lens}}$   & $5.0\times10^{6}M_{\odot}$ & $4.98\times10^{6}M_{\odot}$ & $5.05\times10^{6}M_{\odot}$ \\
        $t_{\text{orbit}}$\,[yr]       & 2078 & 4050 & 8131 \\
        \hline 
    \end{tabular}}
  \tablefoot{The parameters $d$, $v$, $m_{\text{lens}}$, and $M_{\text{center}}$ are given for the maximum magnification of $A = 6.3$.  The scenarios cover the distance parameter space of $160\,\text{ld} \leq d \leq 400\,\text{ld}$ and, correspondingly, the velocity parameter space of $400\,\text{km}\,\text{s}^{-1} \geq v \geq 250\,\text{km}\,\text{s}^{-1}$ for the lensing object.
}      
\label{tab:spectroscopy_log}
\end{table}
In a second step, we expanded the parameter space to distances of 110 and 1000 light days as well as to higher central BH masses of 
$5\times10^7\,M_{\odot}$ and $5\times10^8\,M_{\odot}$ (see Appendix section: Table~\ref{tab:spectroscopy2_log}, Table~\ref{tab:spectroscopy3_log}, Table~\ref{tab:spectroscopy4_log}). 
This leads to orbiting times from 100 to 30\,000 years.

A nearly edge-on binary orientation is necessary to produce self-lensing 
systems of MBHBs, that is, with a sufficiently close angular separation \citep[]{dorazio18, kelley21}.
Light curves of lensing objects are expected to be achromatic. However, in misaligned configurations -- that is to say if the disk plane of the active nucleus does not coincide with the orbital plane of the lens -- the bluer optical bands can be 
lensed more than the redder ones \citep[]{kelley21}. This is what we see in the declining phase based on UV/optical SWIFT observations \citep[]{oknyansky20}.  Furthermore, the difference in the observed amplitudes might be caused, at least in part, by a more significant contribution from the host galaxy at longer wavelengths \citep[]{oknyansky20}.
Light curves for extended accretion disks can be broader and flatter \citep[]{dorazio18} compared to light curves of a lensed point source, as has been described before \citep[]{paczynski91}.
In addition, deflections and time delay effects may change the
size and shape of the self-lensing flares from SMBH binaries \citep[]{davelaar22a}. The self-lensing flares
may lead to additional features in the observed light curve imprinted by the relativistic shadow around the background BH \citep[]{davelaar22b}.        

The BH mass of the nucleus of NGC\,1566 has previously been estimated to be $\sim$ $5\times10^{6} M_{\odot}$ based on line widths.
 This means that the central AGN BH and the lensing BH at a distance on the order of 250\,ld have a mass ratio of 10:1. This BH mass ratio is similar to that found for the innermost nuclei (S1 and S2) in the triple nucleus system \citep[]{kollatschny20} NGC\,6240. In that publication, evidence is given that only one nucleus is active in the central inner system of NGC\,6240; however, the distance between the inner nuclei is much larger (200\,pc) than in NGC\,1566.

%**********************************************************************************
%
\section{Discussion}\label{sec:discussion}
%
%**********************************************************************************

\subsection{Probability of gravitational lensing events}

The predicted detection rates for binary AGNs in photometric surveys 
are on the order of
0.1 - 1 percent of AGNs \citep[]{kelley19}. The probability to 
detect SMBH binaries depends - amongst others - on the masses of the BHs and the binary separation. Higher BH masses lead
to shorter orbital periods and therefore higher numbers of possible 
lensing events (see Tables 1 to 4). When comparing our estimated orbital period of 4000 years with the lensing time of 3 - 4 years, we get a detection 
probability for a lensing  effect on the order of 0.1 percent. 
Similar numbers for a detection probability are given in a further work \citep[]{kelley21}
%, graham15
and the references therein.

\subsection{Emission line profile shifts during the outburst supporting the lensing scenario}

Another approach to detect SMBH binaries is
through spectroscopic observations searching for time-dependent offsets of the broad emission lines that are
characteristics of orbital motions of the lensing source in front of the rotating disk
of the primary BH. It has been shown by means of very high spatial resolution observations with GRAVITY (\cite{gravity18}, their Fig.1b) and  \citet[] {gravity21} that there exists a gradient in the velocity of the 
BLR gas perpendicular to the rotation axis.
%\citep[]{liu14, guo19}. Basic estimates for the observability of such %kinematically offset binaries give probabilities of 0.5 to 0.03
%percent \citep[]{kelley21}.

Spectra of NGC\,1566 have been taken with the ESO VLT and MUSE during the 
increasing phase of the outburst in September 2015 and October 2017. An 
additional spectrum was secured with the SALT telescope in July 2018
immediately after the detection of the outburst with INTEGRAL. Additional 
spectra taken with SALT exist for October 2018 
and September 2019. They are  all described in a recent work \citep[]{ochmann24}. The narrow forbidden lines remained constant during the outburst from 2017
to 2019. The broad Balmer lines varied with respect to their intensity  -- 
simultaneously with the continuum variations.
These spectra bridge the gap in the continuum light curve during the 
maximum in 2018. It is intriguing that the
broad line profiles were systematically shifted blueward by a few hundred km\,s$^{-1}$ from 2015 to 2018 (\cite{ochmann24}, Figs. 7,8). 
Such a shift in velocity of a profile is difficult to understand with normal spectral line variability but it qualitatively fits into the scheme of a resolved BLR emitting region that shows a velocity gradient due to rotation.

\subsection{Substructures in the continuum light curve}

In Figs.~\ref{Malte_lensing_fit_quad_single_20240216.pdf} and \ref{Maltelensing_fit_zoom_20240216.pdf}, we see smaller sub-outbursts superimposed on the light curve of the main outburst lasting from 2017 to 2020. 
It is noteworthy that the smaller outbursts at $\pm$70 days and the stronger outbursts at 
$\pm$185 days occurred symmetrically with respect to the maximum of the light curve. Given the mass of the primary BH of $4.5\times10^{6} M_{\odot}$ and the transverse velocity of the secondary BH of 320\,km\,s$^{-1}$, this corresponds to 290 and 770 gravitational radii $r_g$. 
These additional symmetric outbursts might be caused by symmetrical structures in the accretion disk, for example, spiral arms, ring structures, hot spots, or outflowing wind components.
%However, additional strucures might as well caused by caustic %crossings (see a discussion in Neira+20 and references therein).
It has been pointed out \citep[]{dorazio18, mediavilla15} that gravitational lensing might probe the inner structure of AGN accretion disks. Such structures might also be caused by an extended and structured lensing source. However, in this case, it would be more difficult to explain the temporal symmetry of both structures with respect to the maximum of the outburst. 

%{\bf WK: deeper discussion: extra components caused by spiral arms, ring structures, hot spots in acretion disk? etc. (e.g. Ward,C.+2024, ApJ 961)}

%**********************************************************************************
%
\section{Summary}\label{sec:summary}
%
%**********************************************************************************

Here we present evidence for the detection of a sub-parsec SMBH binary in the nuclear region of the AGN NGC\,1566 based on a gravitational lensing event. The broad emission line profiles in this object showed systematical velocity shifts during the lensing outburst. This supports the concept of the existence of a
sub-parsec SMBH binary in NGC\,1566.
In the near future, the \textit{Vera Rubin} Observatory LSST is expected to  detect tens to hundreds more candidates for MBHBs based on their light curves  -- readers can refer to the estimated number of detections of gravitationally lensed  SMBHs in \cite{dorazio18}.
Furthermore, the final states of SMBH binaries are expected to become the loudest sources 
of gravitational waves to be detected with LISA \citep[]{derosa19}.

%\begin{enumerate}
%    \item ...
%    
%    \item ...
%    
%    
%    \item ...
%    
%\end{enumerate}
%

\begin{acknowledgements}
 The authors acknowledge discussions with Martin W. Ochmann.
They thank Malte A. Probst for helpful discussions and computational advice. The authors greatly acknowledge support by the DFG grants KO 857/35-1 and support of the German Aerospace Center (DLR) within the framework of the ``Verbundforschung Astronomie und Astrophysik'' through grant 50OR2305 with funds from the German Federal Ministry for Economic Affairs and Climate Action (BMWK). Research by D.C. is partly supported by grants from the Israeli Science Foundation (2398/19, 1650/23), and by the DFG (CH71-34-3).

\end{acknowledgements}

%% WARNING
%%-------------------------------------------------------------------
%% Please note that we have included the references to the file aa.dem in
%% order to compile it, but we ask you to:
%%
%% - use BibTeX with the regular commands:
%%   \bibliographystyle{aa} % style aa.bst
%%   \bibliography{Yourfile} % your references Yourfile.bib
%%
%% - join the .bib files when you upload your source files
%-------------------------------------------------------------------

\bibliographystyle{aa} % style aa.bst
\bibliography{literature} % your references Yourfile.bib
%\bibliography{test} % your references Yourfile.bib

%\newpage
%**********************************************************************************
%
\begin{appendix}
\section{Calculation of the lensing black hole mass in NGC\,1566  }\label{sec:appendix}
%
%**********************************************************************************

We fitted the observed light curve of NGC\,1566 with a lensing light curve that would be produced by a single point mass - as is described by \cite{paczynski96}.  
The magnification is given in Eqn. \ref{eqn:magnification} (Eqn. 11 in \cite{paczynski96}):  

\begin{equation}
\label{eqn:magnification}
    A = \frac{u^2+2}{u(u^2+4)^{1/2}},
\end{equation}
where $u$ is the time-dependent angular distance between the lensing object and the lensed object in units of the Einstein angle $\Theta_E$. The time dependence of the parameter $u$ relates to the projected motion of the lensing object over the lensed object. It can be translated in terms of the mass and tangential velocity of the lensing object, its distance to the observer, and the distance of lensing and lensed object along the line of sight. This is defined as the characteristic timescale $t_0$ for a lensing event in  Eqn. \ref{eqn:characteristic_timescale} (Eqn.15 in \cite{paczynski96}): 

\begin{equation}
\label{eqn:characteristic_timescale}
    t_0 = 0.214\,\text{yr}\, \left( \frac{M}{M_{\odot}} \right)^{1/2} \left( \frac{D_d}{10\,\text{kpc}} \right)^{1/2} \left(1 - \frac{D_d}{D_s} \right)^{1/2} \left(\frac{200\,\text{km}\,\text{s}^{-1}}{v} \right)
.\end{equation}
The time dependence of the variable $u$ is given in Eqn. \ref{eqn:angular_distance} (Eqn. 16 in \cite{paczynski96}, where $t_{max}$ is the time of the highest magnification. The dimensionless impact parameter $p$ is defined by the minimum angular distance between the lensing and the lensed object at $t_{max}$\,:

\begin{equation}
\label{eqn:angular_distance}
    u = \left[ p^2 + \left(\frac{t-t_{max}}{t_0} \right)^2 \right]^{1/2}
.\end{equation}

The parameters $p$ and $t_0$ can be measured directly from the light curve: The impact parameter can be deduced from the maximum magnification from Eqns. \ref{eqn:magnification} and \ref{eqn:angular_distance} with $A(t_{max}) = p^2+2 / p(p^2+4)^{1/2}$. With known $p$, the characteristic timescale is evaluated from the magnification at different ratios of $\frac{t-t_{max}}{t_0}$. \\
From the symmetry of the light curve, we found $t_{max} = 58279.8$ in MJD. We tested multiple light curves with maximum magnifications in the interval of 18 to 33 mJy (i.e., fluxes with previously subtracted host contribution) using the minimum $\chi^2$ method. For each tested light curve, the impact parameter $p$ and the first epoch after the peak flux with a magnification falling below $A(\frac{t-t_{max}}{t_0}=0.5)$ was determined. The parameters of our optimal light curve are $p = 0.16$ and $t_0 = 155\pm20\,$days.

Table~\ref{tab:spectroscopy2_log}  in the Appendix section is an extension of Table~\ref{tab:spectroscopy_log}.   We calculated the transverse velocity as well as the masses of the lensing and the lensed BH for two additional scenarios with distances of 110 and 1000 light days, respectively.

Table~\ref{tab:spectroscopy3_log} and Table~\ref{tab:spectroscopy4_log} are extensions of Table~\ref{tab:spectroscopy2_log}. 
We expanded the parameter space because both previously used methods to estimate the BH mass might underestimate the mass. First, the low widths of the broad emission lines$^{18}$ may be due to a face-on view of the accretion disk. Second, the general correlation between central BH mass and absorption line
strength might not be valid for SMBH binaries.
Therefore, we calculated the parameters assuming combined BH masses of $5\times10^7\,M_{\odot}$ in Table~\ref{tab:spectroscopy3_log} and $5\times10^8\,M_{\odot}$ in Table~\ref{tab:spectroscopy4_log} instead of $5\times10^6\,M_{\odot}$.

\begin{table*}[ht!]
\caption{Parameters $d$, $v$, $m_{\text{lens}}$ and $M_{\text{center}}$ for five possible lensing scenarios given the maximum magnification of $A = 6.3$ and the combined mass of the two BHs $M_{\text{center}}+m_{\text{lens}} \approx 5\times10^6\,M_{\odot}$. }
\centering
    \begin{tabular}{lccccc}
        \hline \hline
        \noalign{\smallskip}
         & Scenario 0 & Scenario 1 & Scenario 2 & Scenario 3 & Scenario 4\\
        \noalign{\smallskip}
        \hline 
        \noalign{\smallskip}
        $d$\,[ld]             & 110 & 160 & 250 & 400 & 1000\\
        $v$\,[km\,s$^{-1}$]   & 480 & 400 & 320 & 255 &  160\\
        $m_{\text{lens}}$     & $2.5\times10^{6}M_{\odot}$  & $1.2\times10^{6}M_{\odot}$ & $4.8\,\,\,\,\times10^{5}M_{\odot}$ & $1.9\,\,\,\,\times10^{5}M_{\odot}$ & $3\times10^{4}M_{\odot}$ \\
        $M_{\text{center}}$   & $2.5\times10^{6}M_{\odot}$ & $3.8\times10^{6}M_{\odot}$ & $4.5\,\,\,\,\times10^{6}M_{\odot}$ & $4.86\times10^{6}M_{\odot}$ & $5.0\times10^{6}M_{\odot}$ \\
        $M_{\text{center}}+m_{\text{lens}}$  & $5.0\times10^{6}M_{\odot}$ & $5.0\times10^{6}M_{\odot}$ & $4.98\times10^{6}M_{\odot}$ & $5.05\times10^{6}M_{\odot}$ & $5.03\times10^{6}M_{\odot}$\\
        $t_{\text{orbit}}$\,[yr]    & 1178  & 2050 & 4050 & 8131 & 32155\\
        \hline 
    \end{tabular}
  \tablefoot{The scenarios cover the distance parameter space of $110\,\text{ld} \leq d \leq 1000\,\text{ld}$ and, correspondingly, the velocity parameter space of $480\,\text{km}\,\text{s}^{-1} \geq v \geq 160\,\text{km}\,\text{s}^{-1}$ for the lensing object.
  }
  \label{tab:spectroscopy2_log}
\end{table*}

\begin{table*}[ht!]
\caption{Parameters $d$, $v$, $m_{\text{lens}}$ and $M_{\text{center}}$ for three possible lensing scenarios given the maximum magnification of $A = 6.3$ and the combined mass of the two BHs $M_{\text{center}}+m_{\text{lens}} \approx 5\times10^7\,M_{\odot}$. }

\centering
    \begin{tabular}{lccccc}
        \hline \hline
        \noalign{\smallskip}
         & Scenario 0 & Scenario 1 & Scenario 2 & Scenario 3 & Scenario 4\\
        \noalign{\smallskip}
        \hline 
        \noalign{\smallskip}
        $d$\,[ld]             & 110 & 160 & 250 & 400 & 1000\\
        $v$\,[km\,s$^{-1}$]   & 1530 & 1270 & 1010 & 805 &  510\\
        $m_{\text{lens}}$     & $2.5\times10^{7}M_{\odot}$  & $1.2\times10^{7}M_{\odot}$ & $4.8\,\,\,\,\times10^{6}M_{\odot}$ & $1.9\,\,\,\,\times10^{6}M_{\odot}$ & $3\times10^{5}M_{\odot}$ \\
        $M_{\text{center}}$   & $2.5\times10^{7}M_{\odot}$ & $3.8\times10^{7}M_{\odot}$ & $4.5\,\,\,\,\times10^{7}M_{\odot}$ & $4.86\times10^{7}M_{\odot}$ & $5.0\times10^{7}M_{\odot}$ \\
        $M_{\text{center}}+m_{\text{lens}}$  & $5.0\times10^{7}M_{\odot}$ & $5.0\times10^{7}M_{\odot}$ & $4.98\times10^{7}M_{\odot}$ & $5.05\times10^{7}M_{\odot}$ & $5.03\times10^{7}M_{\odot}$\\
        $t_{\text{orbit}}$\,[yr]    & 372  & 648 & 1279 & 2571 & 10168\\
        \hline 
    \end{tabular}
    \tablefoot{The scenarios cover the distance parameter space of $110\,\text{ld} \leq d \leq 1000\,\text{ld}$ and, correspondingly, the velocity parameter space of $1530\,\text{km}\,\text{s}^{-1} \geq v \geq 510\,\text{km}\,\text{s}^{-1}$ for the lensing object.
    }
\label{tab:spectroscopy3_log}
\end{table*}
  \begin{table*}[ht!]
\caption{Parameters $d$, $v$, $m_{\text{lens}}$ and $M_{\text{center}}$ for five possible lensing scenarios given the maximum magnification of $A = 6.3$ and the combined mass of the two BHs $M_{\text{center}}+m_{\text{lens}} \approx 5\times10^8\,M_{\odot}$. }
\centering
    \begin{tabular}{lccccc}
        \hline \hline
        \noalign{\smallskip}
         & Scenario 0 & Scenario 1 & Scenario 2 & Scenario 3 & Scenario 4\\
        \noalign{\smallskip}
        \hline 
        \noalign{\smallskip}
        $d$\,[ld]             & 110 & 160 & 250 & 400 & 1000\\
        $v$\,[km\,s$^{-1}$]   & 4840 & 4020 & 3200 & 2550 &  1610\\
        $m_{\text{lens}}$     & $2.5\times10^{8}M_{\odot}$  & $1.2\times10^{8}M_{\odot}$ & $4.8\,\,\,\,\times10^{7}M_{\odot}$ & $1.9\,\,\,\,\times10^{7}M_{\odot}$ & $3\times10^{6}M_{\odot}$ \\
        $M_{\text{center}}$   & $2.5\times10^{8}M_{\odot}$ & $3.8\times10^{8}M_{\odot}$ & $4.5\,\,\,\,\times10^{8}M_{\odot}$ & $4.86\times10^{8}M_{\odot}$ & $5.0\times10^{8}M_{\odot}$ \\
        $M_{\text{center}}+m_{\text{lens}}$  & $5.0\times10^{8}M_{\odot}$ & $5.0\times10^{8}M_{\odot}$ & $4.98\times10^{8}M_{\odot}$ & $5.05\times10^{8}M_{\odot}$ & $5.03\times10^{8}M_{\odot}$\\
        $t_{\text{orbit}}$\,[yr]    & 118  & 205 & 405 & 813 & 3216\\
        \hline 
    \end{tabular}
     \tablefoot{The scenarios cover the distance parameter space of $110\,\text{ld} \leq d \leq 1000\,\text{ld}$ and, correspondingly, the velocity parameter space of $4840\,\text{km}\,\text{s}^{-1} \geq v \geq 1600\,\text{km}\,\text{s}^{-1}$ for the lensing object.
     }
\label{tab:spectroscopy4_log}
\end{table*}

\end{appendix}
\end{document}